\documentstyle[preprint,pre,aps]{revtex}

\newcommand{\gb}{Gay-Berne}

\newcommand{\disc}{disclination}

\newcommand{\nhat}{\mathbf{\hat{n}}}

\newcommand{\beqn}{\protect\begin{equation}}
\newcommand{\eeqn}{\protect\end{equation}}
\begin{document}
\draft
\title{Defect configurations and dynamical behavior in a Gay-Berne nematic
emulsion}
\author{Jeffrey L. Billeter and Robert A. Pelcovits}
\address{Department of Physics, Brown University, Providence, RI 02912}
\date{\today}
\maketitle
\begin{abstract}
To model a nematic emulsion consisting of a surfactant--coated water droplet dispersed in a nematic host, we performed a molecular dynamics simulation of a droplet immersed in
a system of 2048 Gay-Berne ellipsoids in a nematic phase. Strong radial anchoring
at the surface of the droplet induced a Saturn ring defect configuration,
consistent with theoretical predictions for very small droplets. A surface
ring configuration was observed for lower radial anchoring strengths, and a pair of point defects was found near the poles of the droplet for tangential anchoring. We also
simulated the falling ball experiment and measured the drag force anisotropy, in the presence of strong radial anchoring as well as zero anchoring strength.
\end{abstract}
\pagestyle{headings}
\section{Introduction}
With a wide variety of practical applications and numerous opportunities for
investigating basic questions in chemistry and physics, colloidal systems (suspensions of solid particles) and emulsions (dispersions of surfactant-coated liquid droplets) are
of great interest to a variety of researchers. Recently, there has
been increased attention given to emulsions involving water droplets dispersed in liquid crystals \cite{poulin}. 
Surfactant-coated water droplets dispersed in a nematic host (so-called inverted emulsions, as opposed to direct emulsions where the nematic itself is separated into spherical droplets) can serve as generators of topological defects because of the anchoring of the liquid crystalline molecules to the droplet surfaces. Sufficiently strong homeotropic anchoring at the surface of the droplet induces a radial hedgehog defect
centered on the droplet. To compensate for this induced defect, another defect must
be created in the medium. 
\par
There are two defect configurations which are 
theorized to arise in this case---the hyperbolic dipole \cite{lub-pettey} and the Saturn ring \cite{terentjev1} (see Fig. \ref{satdipfig}). 
In the dipole case, the radial hedgehog defect is
compensated by another point defect (a hyperbolic hedgehog); the two together
form the dipole. The Saturn ring is a hyperbolic wedge disclination line which
encircles the droplet equatorially. At large distances, the Saturn ring has the
appearance of a hyperbolic hedgehog. In both cases then, the radial hedgehog with
defect index $+1$ is compensated by a hyperbolic defect with index $-1$. Note that
in the dipole case, the line joining the droplet and the hyperbolic monopole is
parallel to the director $\mathbf{\hat{n}}$, while in the Saturn ring case, the normal
to the plane of the ring is parallel to $\mathbf{\hat{n}}$. Theoretical and numerical work based on elastic theory \cite{lub-pettey,terentjev1,stark1,ruh-ter2} suggests that the dipole configuration is stable for micron sized droplets and sufficiently strong surface anchoring, while a Saturn ring should appear if the droplet size is reduced.
If the strength of the surface anchoring is reduced a third director configuration can arise, termed the surface-ring configuration  \cite{ruh-ter1,stark1,kuksenok}. 
In this configuration the molecules at the equator are
actually parallel to the droplet's surface (homotropic), and radial anchoring is limited
to the poles of the droplet.  Experiments to-date have always revealed dipolar 
configurations \cite{poulin},
consistent with the relatively large particle sizes and surface anchoring used. Changing the surfactant compound to produce homotropic (tangential) anchoring at the surface results in a pair of point
defects at the poles of the droplet (boojums).  
\par
The dynamics of the dispersed droplets and their accompanying director configurations are also of interest \cite{ruh-ter1,ruh-ter4}. Detailed knowledge of the flow properties of liquid crystals is important for both fundamental and technological reasons. One way to measure viscosities in a fluid is via the ``falling ball experiment''. The drag force on a falling ball can be related to the viscosity by a Stokes-type formula: $F_d = - 6 \pi R v \eta$, where $R$ is the radius of the ball, $v$ is its terminal velocity and $\eta$ is the fluid viscosity. Because of the anisotropy of liquid crystalline molecules the drag force on a moving water droplet in an inverse nematic emulsion will depend on the relative orientation of the velocity and the director, i.e., a Stokes-like formula incorporates a nematic resistance tensor $M$\cite{ruh-ter1,ruh-ter4}: 
\beqn
{\bf F} = M_\perp {\bf v} + (M_\parallel -M_\perp) (\bf{v \cdot \nhat})\nhat,
\label{force}
\eeqn
where the subscripts $\perp$ and $\parallel$ refer respectively to directions perpendicular and parallel to the director.
Eqn (\ref{force}) implies that the drag force will not be antiparallel to the velocity, but will instead have a component, $F_{lift}$ perpendicular to $\bf v$. 
The drag force is particularly sensitive to the defect structure around the droplet because regions of high gradients in the director field (such as those around defects) result in higher resistance to flow. Theoretical work on the drag force exerted on spheres as well as cylinders has been carried out by Ruhwandl and Terentjev \cite{ruh-ter1,ruh-ter4} in the low Ericksen number regime (small values of $vR\eta/K$ where $v$ and $R$ are the droplet velocity and radius respectively and $\eta$ and $K$ are characteristic values of the nematic viscosity and elastic constant respectively) where they found a drag force anisotropy (the ratio $F_\perp/F_\parallel$) in the range of 1.5-1.75. A drag force anisotropy not equal to unity is equivalent to a nonzero lift force. These authors also considered discotic molecules \cite{ruh-ter4} and found a ratio less than unity which is reasonable given the oblate shape of the molecule.
\par
Here we present the results of molecular dynamics (MD) simulations carried out on a system consisting of one spherical droplet immersed in a \gb\ nematic liquid crystal \cite{gayberne}. The \gb\ model is a phenomenological single site model of mesogenic molecules which captures the essential physical features of a wide variety of liquid crystalline behavior \cite{bates}. Using a fluid model like the \gb\ system, as opposed to a lattice model of liquid crystals, we are able to simulate not only static behavior such as the director configuration surrounding the sphere (including the role of thermal fluctuations which are neglected in elastic theory) but dynamical behavior as well. We model the interaction of the \gb\ molecules with the droplet using a generalized repulsive \gb\ potential \cite{cleaver-care} to account for excluded volume effects along with a phenomenological surface anchoring potential. Due to the large computational overhead of the \gb\ system our study is restricted to  2048 \gb\ particles and a relatively small droplet, with a diameter equal to the long axis of the \gb\ molecule. However, using massively parallel computers it is possible to study much larger \gb\ systems consisting of at least 65,000 molecules \cite{wilson,billeter} and we hope to carry out a simulation of the emulsion system with a similar number of host molecules in the future. 
\par
The outline of this paper is as follows. In the next section we present the computational details of our study. Section III contains the results for the director configuration in the absence of an external force on the droplet, as well as the results of ``falling ball'' type experiments. We offer some concluding remarks in Section IV.
\section{Computational Details}
We model the nematic host using the original \gb\ potential given by \cite{luck1}:
\begin{displaymath} U\left({\bf \hat{u}_i,\hat{u}_j,\hat{r}}\right)=4\varepsilon
\left({\bf \hat{u}_i,\hat{u}_j,\hat{r}}\right) \end{displaymath} \begin{equation} 
\times\left[\left\{\frac{\sigma_0}{r-\sigma\left({\bf \hat{u}_i,\hat{u}_j,\hat{r}}
\right)+\sigma_0}\right\}^{12}\!-\left\{\frac{\sigma_0}
{r-\sigma\left({\bf \hat{u}_i,\hat{u}_j,\hat{r}}\right)+\sigma_0}\right\}^6\right],
\label{GBpot}\end{equation} where ${\bf \hat{u}_i,\hat{u}_j}$
give the orientations of the long axes of molecules $i$ and $j$, respectively, 
and ${\bf r}$ is the intermolecular
vector (${\bf r}={\bf r_i}-{\bf r_j}$).
The parameter $\sigma\left({\bf \hat{u}_i,\hat{u}_j,\hat{r}}\right)$ is the 
intermolecular separation at which the potential vanishes, and thus represents the
shape of the molecules. Its explicit form is
\begin{eqnarray} \sigma\left({\bf \hat{u}_i,\hat{u}_j,\hat{r}}\right)&=&
\sigma_0\left[1-\frac{1}{2}\chi\left\{
\frac{\left({\bf \hat{r}\cdot\hat{u}_i}+{\bf \hat{r}\cdot\hat{u}_j}\right)^2}{
1+\chi\left({\bf \hat{u}_i\cdot\hat{u}_j}\right)}\right.\right. \nonumber\\ 
&&\left.\left.{}+
\frac{\left({\bf \hat{r}\cdot\hat{u}_i}-{\bf \hat{r}\cdot\hat{u}_j}\right)^2}{
1-\chi\left({\bf \hat{u}_i\cdot\hat{u}_j}\right)}\right\}\right]^{-1/2}, \label{sigma}\end{eqnarray}
where $\sigma_0=\sigma_s$ (defined below) and $\chi$ is
\begin{equation} \chi=\left\{\left(\sigma_e/\sigma_s\right)^2-1\right\}/
\left\{\left(\sigma_e/\sigma_s\right)^2+1\right\}. \end{equation} Here $\sigma_e$
is the separation between two molecules when they are oriented end-to-end, and
$\sigma_s$ the separation when side-by-side. 
The well depth $\varepsilon\left({\bf \hat{u}_i,\hat{u}_j,\hat{r}}\right)$,
representing the anisotropy of the attractive interactions, is
written as \begin{equation}\varepsilon\left({\bf \hat{u}_i,\hat{u}_j,\hat{r}}\right)
=\varepsilon_0\varepsilon^\nu\left({\bf \hat{u}_i,\hat{u}_j}\right)\varepsilon'^{\mu}
\left({\bf \hat{u}_i,\hat{u}_j,\hat{r}}\right),\end{equation} where
\begin{equation}\varepsilon\left({\bf \hat{u}_i,\hat{u}_j}\right)=\left\{1-
\chi^2\left({\bf \hat{u}_i\cdot\hat{u}_j}\right)^2\right\}^{-1/2},\label{eps}\end{equation} and
\begin{eqnarray} \varepsilon'\left({\bf \hat{u}_i,\hat{u}_j,\hat{r}}\right)&=&
1-\frac{1}{2}\chi'\left\{
\frac{\left({\bf \hat{r}\cdot\hat{u}_i}+{\bf \hat{r}\cdot\hat{u}_j}\right)^2}{
1+\chi'\left({\bf \hat{u}_i\cdot\hat{u}_j}\right)}\right. \nonumber\\ 
&&\left.{}+
\frac{\left({\bf \hat{r}\cdot\hat{u}_i}-{\bf \hat{r}\cdot\hat{u}_j}\right)^2}{
1-\chi'\left({\bf \hat{u}_i\cdot\hat{u}_j}\right)}\right\}, \label{epsprime}\end{eqnarray}
with $\chi'$ defined in terms of $\varepsilon_e$ and $\varepsilon_s$, the end-to-end 
and side-by-side well depths, respectively, as
\begin{equation} \chi'=\left\{1-\left(\varepsilon_e/\varepsilon_s\right)^{1/\mu}\right\}/
\left\{1+\left(\varepsilon_e/\varepsilon_s\right)^{1/\mu}\right\}. \end{equation}
The overall energy scale is set by the value of $\varepsilon_0$. \par
For the adjustable parameters, we primarily used the values originally proposed by Gay and Berne \cite{gayberne}:  $\mu=2,\nu=1,\sigma_e/\sigma_s=3$, and
$\varepsilon_s/\varepsilon_e=5$. We also ran simulations with the parameterization suggested by Berardi et al. \cite{ber-emer} (where $\mu=1$ and $\nu=3$) and saw no noticeable differences in our results.
 \par
The interaction between the \gb\ molecules and a droplet of radius $R$ consists of
two parts---a soft, repulsive term $U_r$ and an anchoring term $U_a$.  The repulsive term utilizes the parameters of
a \gb\ potential generalized  to mimic the interaction between nonequivalent particles \cite{cleaver-care}.  The  range parameter $\sigma_{r}\left(\mathbf{\hat{u}, \hat{r}}
\right)$ and the energy parameter $\varepsilon_{r}\left(\mathbf{\hat{u}}\right)$
used in this potential are generalizations of the corresponding parameters in Eqns. (\ref{sigma})
and (\ref{eps}), namely \beqn \sigma_{r}\left({\bf \hat{u},\hat{r}}\right)=\sigma_o^{r}\left(1-\chi_{r}\left(
\bf \hat{r}\cdot\hat{u}\right)\right)^{-1/2} \eeqn
and
\beqn \varepsilon_{r}\left({\bf \hat{u},\hat{r}}\right) = 1,\eeqn
with  $\chi_{r}$ defined as 
\beqn \chi_{r}=\frac{\sigma_e^2-\sigma_s^{2}}
{\sigma_e^2+4R^2} ,\eeqn
and 
\beqn \sigma_o^{r}=\frac{1}{\sqrt{2}}(\sigma_s^2 + 4R^2)^{1/2}.\eeqn
 Thus the repulsive part of the potential between a molecule with orientation $\bf \hat{u}$ and the droplet with center of mass separation $\bf r$ is given by
\beqn U_{r}\left({\bf \hat{u},\hat{r}}\right)=4\varepsilon_0  
\times\left\{\frac{\sigma_0}{r-\sigma_{r}\left({\bf \hat{u},\hat{r}}
\right)+\sigma_0}\right\}^{18}
\label{GBpot_rd}.\eeqn
 We use an exponent of $18$ instead of $12$ in this repulsive term to make
the interaction ``harder''.
\par
The short-ranged surface anchoring term is given by: 
\beqn U_{a}=-W\frac{(\mathbf{\hat{r}} \cdot \mathbf{\hat{u}})^6}{r^{6}},
\label{anchterm}\eeqn where $W$ is a phenomenological anchoring coefficient. Only those molecules with $r < R+\sigma_e$ are
considered to interact with the droplet via this surface anchoring. A positive value of $W$ yields homeotropic anchoring, while a negative
value of $W$ gives tangential anchoring. 
  \par
We performed  MD simulations at constant temperature, $T^*\equiv k_BT/\varepsilon_o=0.95$, and pressure, $P^\ast=P\sigma_o^3/\varepsilon_o=4.0$, corresponding to a nematic phase. We controlled temperature and pressure using 
 the equations of motion of refs. \cite{melchi} and \cite{bulg-adam}, which are modifications of the original Nose-Hoover approach \cite{nose}.  The orientational degrees of freedom were parameterized using quaternions.
We used a variation \cite{svanberg} of the leapfrog algorithm to solve these equations for a system consisting of 2048 \gb\ molecules and one spherical droplet with a timestep $\Delta t^* = 0.001, \Delta t^* \equiv (m \sigma_o^2/\varepsilon_o)^{-1/2}$, where $m$ is the mass of the \gb\ molecule. The droplet diameter was chosen to be $3\sigma_0$ which is the length of the long axis of the \gb\ molecules in the host, and the mass of the droplet was chosen to be $100m$, so that in the absence of an external force the sphere moves very slowly compared to
 the molecules.  The surface anchoring parameter $W$ was chosen to be 4000 (in dimensionless units) for weak anchoring and 35000 for strong anchoring. In terms of the dimensionless quantity $\xi=K/WR$ which measures 
the elastic deformation energy relative to the surface anchoring energy \cite{degennes}, the former value of $W$ corresponds to $\xi \approx 0.4$, while the latter corresponds to a value of $\xi$ ten times smaller (these values assume that $K$ is of order unity). We prepared our system by placing the sphere at the center of an empty MD cell. We then surrounded the sphere with eight replicas of a \gb\ nematic phase of 256 particles, equilibrated to the temperature and pressure specified above. The system was then equilibrated for 100000 timesteps.
\section{Results}
\subsection{Static Director Configuration}
  
In the absence of an external force on the droplet we studied the director configuration using a variety of techniques and considered both strong and weak surface anchoring. After the equilibration process described in the previous section we carried out production runs of 60000 steps examining the director configuration every 6000 steps. The general features of the configurations required to assess the overall director structure were essentially time-independent. We studied these configurations using the defect-finding algorithm for disclinations described in \cite{zap-gold2}, and contour plots of the angle of molecular orientation with respect to the director.
To apply the defect-finding algorithm we divided the system into a $6 \times 6 \times 6$ lattice of cubic bins (the size is chosen so that the defects ultimately found have a core size of one bin).
Note that the creation of a lattice is strictly for convenience in defect finding; the time
evolution of the system allows for complete translational freedom.  Within each bin, the order parameter
tensor \begin{equation} \label{ordparamtens} 
Q_{\alpha\beta}({\bf x}) = \frac{3}{2}\left[
{\bf \hat{u}}_\alpha({\bf x}){\bf \hat{u}}_\beta({\bf x}) - 
\delta_{\alpha\beta}\right],
\end{equation}  was calculated, its largest eigenvalue taken as the local
order parameter $S$ and the corresponding eigenvalue as the local director ${\bf \hat{n}}$.
Topological defects can be located by considering the directors at the corners of a square (one of the faces of a cube) in 
the 3D lattice. The idea is to track the course of the intersections of these
directors with the order parameter sphere (actually the projective plane $RP_2$) as one
moves around the corners of the real-space square (Fig. \ref{discalgfig}). Starting with the
intersection of ${\bf \hat{n}}_A$ with the sphere, one then takes as the next point either
the intersection of ${\bf \hat{n}}_B$ or $-{\bf \hat{n}}_B$, whichever is closest 
to ${\bf \hat{n}}_A$'s
intersection. Once this point is determined, either ${\bf \hat{n}}_C$ or 
$-{\bf \hat{n}}_C$ is
used, depending on the proximity to the previously defined point, and so on. Once the last
point (from corner D) is determined, one looks at whether its intersection is in the same
hemisphere as the starting point's. If so, no defect is present---the path in order
parameter space is deformable to a single point, \textit{i.e.,} a uniform configuration.
If the first and last points are in different hemispheres, however, then a \disc\ line
is taken to cut through the center of the square and is oriented perpendicular to the plane of 
the square.
\par
For strong anchoring ($W=35000, \xi=0.04$) we detected a  Saturn ring configuration using this algorithm (Fig. \ref{sphsatdiscfig}). Given the small size of our droplet this result is  consistent with the results of elastic theory \cite{lub-pettey,terentjev1,stark1,ruh-ter2}.  The molecular configuration in the portion of the MD cell surrounding the droplet is shown in Fig. \ref{sphsatvecfig} providing a clear view of the strong radial anchoring of the molecules immediately surrounding the sphere. It is clear that the radial anchoring takes place over one molecular layer. Thus, it
is not possible with our system size to provide more than an estimate of the radius of the Saturn ring, namely, it is slightly greater than the radius of the droplet. However, our result  is consistent with the results of elastic theory analyses of 
references \cite{stark1} and \cite{ruh-ter3}, who find radii of order $R$. With a
larger system size and thus a larger droplet, it should be possible to determine the 
radius of the ring. 

\par
The director configurations can also be visualized using contour plots of the angle of orientation with respect to the director. As
shown in Fig. \ref{basicparslfig}
(a slice though the center of the droplet and perpendicular to 
$\mathbf{\hat n}$), the molecular orientation is along the director everywhere except
in a ring about the droplet. Fig. \ref{basicperpslfig} shows a slice 
parallel to $\mathbf{\hat n}$. Both figures are consistent with the presence of a Saturn ring in the equatorial plane perpendicular to $\mathbf{\hat n}$.
A similar set of figures for tangential anchoring (Figs. \ref{basicparfig} and \ref{basicpar2fig})
show a pair of boojums located near the poles of the droplet. 
\par
For the lower value of the anchoring parameter that we studied, $W=4000$, we found that the Saturn ring is replaced by the surface-ring configuration as shown in Fig. \ref{surfvecfig}. This value of $W$ corresponds to a surface extrapolation length $\xi= 0.4$, and our result is  consistent with the predictions of elastic theory \cite{stark1,ruh-ter3}.
\par
\subsection{Dynamics}
To study the dynamics of the immersed droplet we began with an equilibrated configuration prepared as described at the end of section II. We applied a force to the droplet and measured its position as a function of time, determining the terminal velocity from the asymptotic slope of the droplet's position as a function of time (see Fig. \ref{f85}). For a force applied at an angle of $45^\circ$ with the director the lift force ratio (the ratio of the components of drag force perpendicular and parallel to the velocity) is given by,

\beqn
F_{lift} / F_v = \tan (45^\circ-\theta),
\eeqn
where $\theta$ is the angle between the terminal velocity vector and the director. The anisotropy of the resistance tensor $M$ is given by,
\beqn
M_\perp /M_\parallel = v_\parallel/ v_\perp = \cot\theta.
\eeqn
Table \ref{table1} shows the magnitude of the terminal velocity, the value of $\theta$, the lift force ratio and the resistance tensor anisotropy for driving forces of different magnitudes for both
$W=35000$ (strong anchoring) and for $W=0$. 
The magnitudes of $M_\perp$ and $M_\parallel$ in the former case are approximately five times greater than in the latter, as one might expect intuitively. For both cases, the values of $\theta$ are generally slightly less than $45^\circ$. Thus, the lift force ratio is nearly zero, and the resistance anisotropy nearly unity. These results are very different from what is found experimentally and predicted theoretically \cite{ruh-ter1,ruh-ter4} for droplets which are many times larger than the nematic host molecules (the theoretical work uses the Frank elastic energy, and thus assumes large droplets). However, given our small droplet size (diameter equal to the length of a nematic molecule), it is not surprising that the lift force $F_{lift}$ is very small, and the resistance tensor nearly isotropic. 
\par
 In the anchoring case, there is
very little distortion of the Saturn rings for small driving forces. With larger
forces, however, the Saturn ring either folds back on itself behind the droplet
for parallel driving forces (Figs. \ref{fpar250parslfig} and \ref{fpar250perpslfig})
or becomes stretched out behind the droplet while retaining its equatorial position
for perpendicular driving forces (Figs. \ref{fperp250parslfig} and \ref{fperp250perpslfig}).
The distortions become so large that, as one can clearly see in Fig. \ref{fpar250perpslfig},
for example, when the droplet exits the left side of the simulation box and re-enters
on the right, it will begin to interact with its own wake. For the no-anchoring case,
there is no Saturn ring to distort, but the droplet still causes localized
disturbances which eventually result in self-interactions for large driving forces.

\section{Conclusions}
We have shown in an off-lattice simulation that, consistent with theoretical
predictions, the Saturn ring is the preferred defect configuration around 
small droplets with radial anchoring embedded in a nematic. We have also
shown that lowering the anchoring strength can produce a surface ring instead
of a Saturn ring. The Saturn ring increases (compared to the no-anchoring case) 
resistance to the droplet's motion when the droplet is subjected to a driving force.
We found a very small lift force (and correspondingly little anisotropy in the resistance tensor), consistent with the small droplet size studied. Clearly, further
work with larger system sizes will allow numerous interesting questions to be
studied fruitfully. The size limitation on the droplet comes, of course, from periodic boundary conditions
; if the droplet is too large, the anchoring layers on opposite sides of the
droplet will begin to interact with each other. Larger system sizes will also allow
exploration of the variation in Saturn ring radius with particle size and, very
importantly, study of the transition from Saturn ring to dipole. With larger
systems, one can also reasonably begin to study multiparticle dispersions.

\section*{Acknowledgments}
We are grateful to Prof. G. Crawford for helpful discussions. Computational work in support of this research was performed at the
Theoretical Physics Computing Facility at Brown University. This work was 
supported by the National Science Foundation under grants 
DMR-9528092 and DMR98-73849.
\bibliographystyle{prsty}

\begin{figure}
\caption{Two possible defect configurations surrounding a water droplet placed in a nematic host, with strong radial surface anchoring. The droplet is indicated by the large sphere. (a) Dipole configuration showing the  induced hyperbolic monopole, indicated by the small sphere.
(b) Saturn ring consisting of an induced wedge disclination ring about the droplet. The Saturn ring
extends into and out of the plane of the page.}
\label{satdipfig}\end{figure}

\begin{figure}
\caption{The disclination-finding algorithm. The directors at the corners of the
lattice cube face shown on the left are tracked on the order parameter space
sphere shown on the right. The diameters $AA^\prime,BB^\prime,CC^\prime,DD^\prime$ correspond to the axes of the headless director at the real--space lattice sites $A,B,C,D$ respectively.}
\label{discalgfig}
\end{figure}
\begin{figure}
\caption{A Saturn ring surrounding the droplet (of diameter $3 \sigma_o$) at the center of the MD cell, with surface anchoring parameter $W=35000$ (strong anchoring). The ring was located using the disclination-finding algorithm
described in \protect\cite{zap-gold2}. The MD cell was divided into a  $6 \times 6 \times 6$ lattice leading to the jagged appearance of the ring. Most of the ring segments are located in the equatorial plane perpendicular to
the director.}
\label{sphsatdiscfig}\end{figure}
\begin{figure}
\caption{Molecular configuration for strong anchoring. A subset of the
total simulation box centered on the droplet is shown and those molecules closest to the droplet are
highlighted to assist in viewing the defect configuration.}
\label{sphsatvecfig}\end{figure}

\begin{figure}
\caption{Contour plot of the director configuration in a slice centered on a droplet 
with strong radial anchoring. The lighter regions represent molecules aligned with the director, 
while darker regions indicate molecules perpendicular to the director. Here the director
points up out of the page, so the figure shows a Saturn ring in the plane of 
the figure.}
\label{basicparslfig}\end{figure}
\begin{figure}
\caption{Contour plot of the director configuration for strong radial anchoring in a slice with the director in the plane of the page as indicated. The lighter regions represent molecules aligned with the director, 
while darker regions indicate molecules perpendicular to the director.}
\label{basicperpslfig}\end{figure}
\begin{figure}
\caption{Contour plot of the director configuration with parallel surface anchoring in a slice perpendicular to the director. Compare with the case of radial anchoring shown in Fig. \ref{basicparslfig}.}
\label{basicparfig}\end{figure}
\begin{figure}
\caption{Contour plot of the director configuration with parallel surface anchoring in a slice in the plane of the director. Compare with the case of radial anchoring shown in Fig. \ref{basicperpslfig}.}
\label{basicpar2fig}\end{figure}
\begin{figure}
\caption{The same as Fig. \ref{sphsatvecfig} but with weaker radial anchoring ($W=4000$).
The Saturn ring has contracted to a surface ring.}
\label{surfvecfig}\end{figure}
\begin{figure}
\caption{Illustration of asymptotic velocity measurement. Sphere positions $x,y,z$ (top to bottom curves respectively) are shown for a driving force $F=85$ applied at $45^\circ$ to the director. The slopes of the straight lines yield the corresponding asymptotic velocity components $v_x,v_y,v_z$.}
\label{f85}
\end{figure}
\begin{figure}
\caption{Slice perpendicular to the director for driving force $F=250,$ with the force applied parallel to the director.}
\label{fpar250parslfig}\end{figure}
\begin{figure}
\caption{Slice in the plane of the director for driving force $F=250,$ with the force applied parallel to the director.
The Saturn ring is clearly distorted.}
\label{fpar250perpslfig}\end{figure}
\begin{figure}
\caption{Slice perpendicular to the director for driving force $F=250,$ with the force applied perpendicular to the director.
Again, the Saturn ring is clearly distorted.}
\label{fperp250parslfig}\end{figure}
\begin{figure}
\caption{Slice in the plane of the director for driving force $F=250,$ with the force applied perpendicular to the director.}
\label{fperp250perpslfig}\end{figure}
\begin{table}
\caption{Lift force ratio and resistance tensor anisotropy. A force of magnitude $F$ is applied at an angle of $45^\circ$ to the director and the magnitude $v$ and angle $\theta$ (with respect to the director) of the terminal velocity are measured. The lift force ratio $F_{lift} / F_v$ is given by $\tan(45^\circ-\theta)$ and the resistance tensor anisotropy $M_\perp /M_\parallel$ by $\cot\theta$.
.} 
\label{table1}
\begin{center}
\begin{tabular}{lcccc}
$F$ & $v$ &$\theta$& $F_{lift} / F_v$ &$M_\perp / M_\parallel$\\ 
\tableline
\multicolumn{5}{c} {$W=35000$}\\
30&0.09&43.5&0.027&1.06\\
50&0.12&43.5&0.026&1.05\\
85&0.17&44.9&0.002&1.00\\
200&0.24&44.8&0.004&1.01\\ 
\tableline
\multicolumn{5}{c} {$W=0$}\\
5&0.06&37.7&0.128&1.29\\
20&0.34&44.5&0.009&1.02\\
40&0.37&42.7&0.041&1.08\\
\end{tabular} 
\end{center}
\end{table}

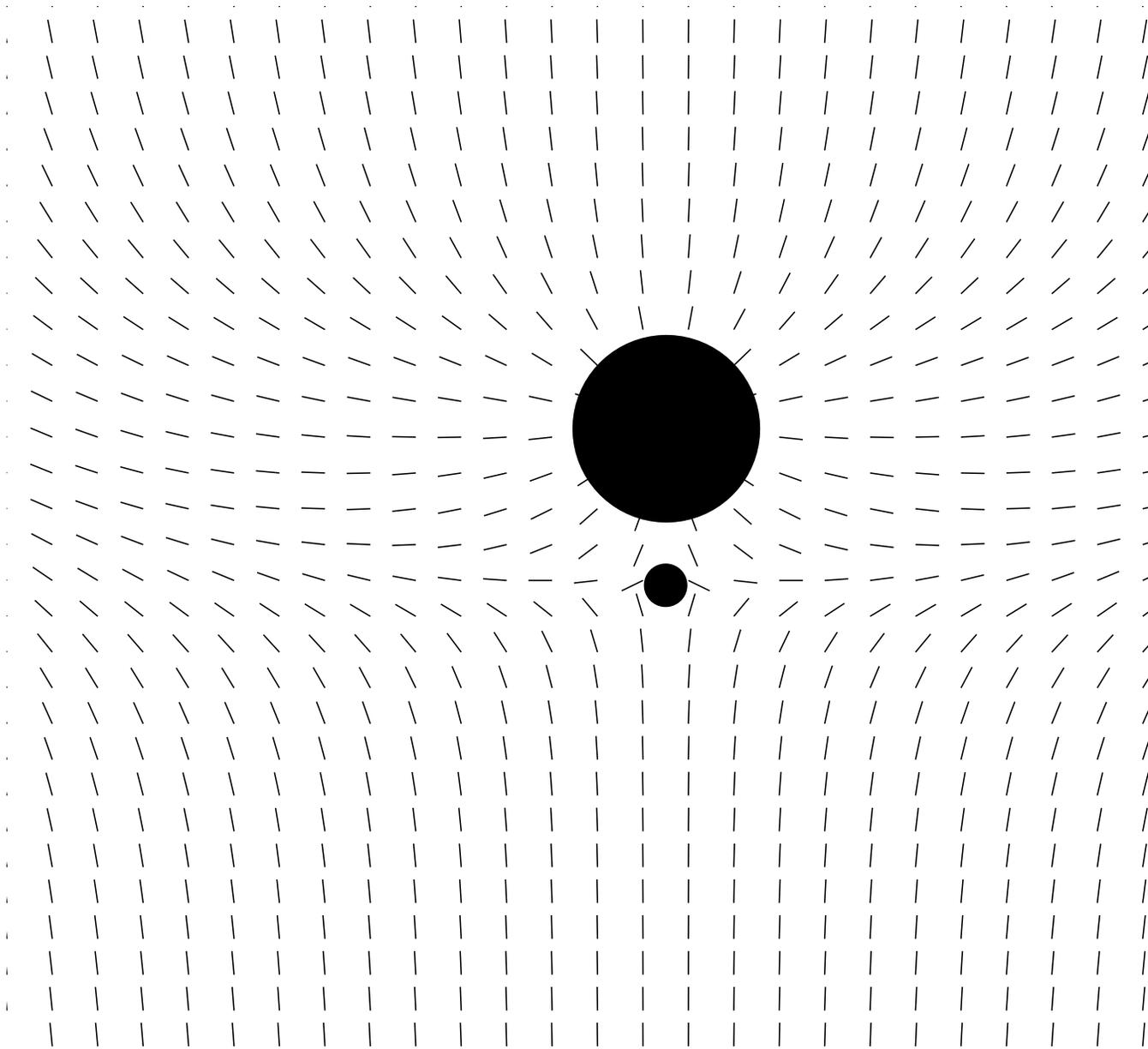
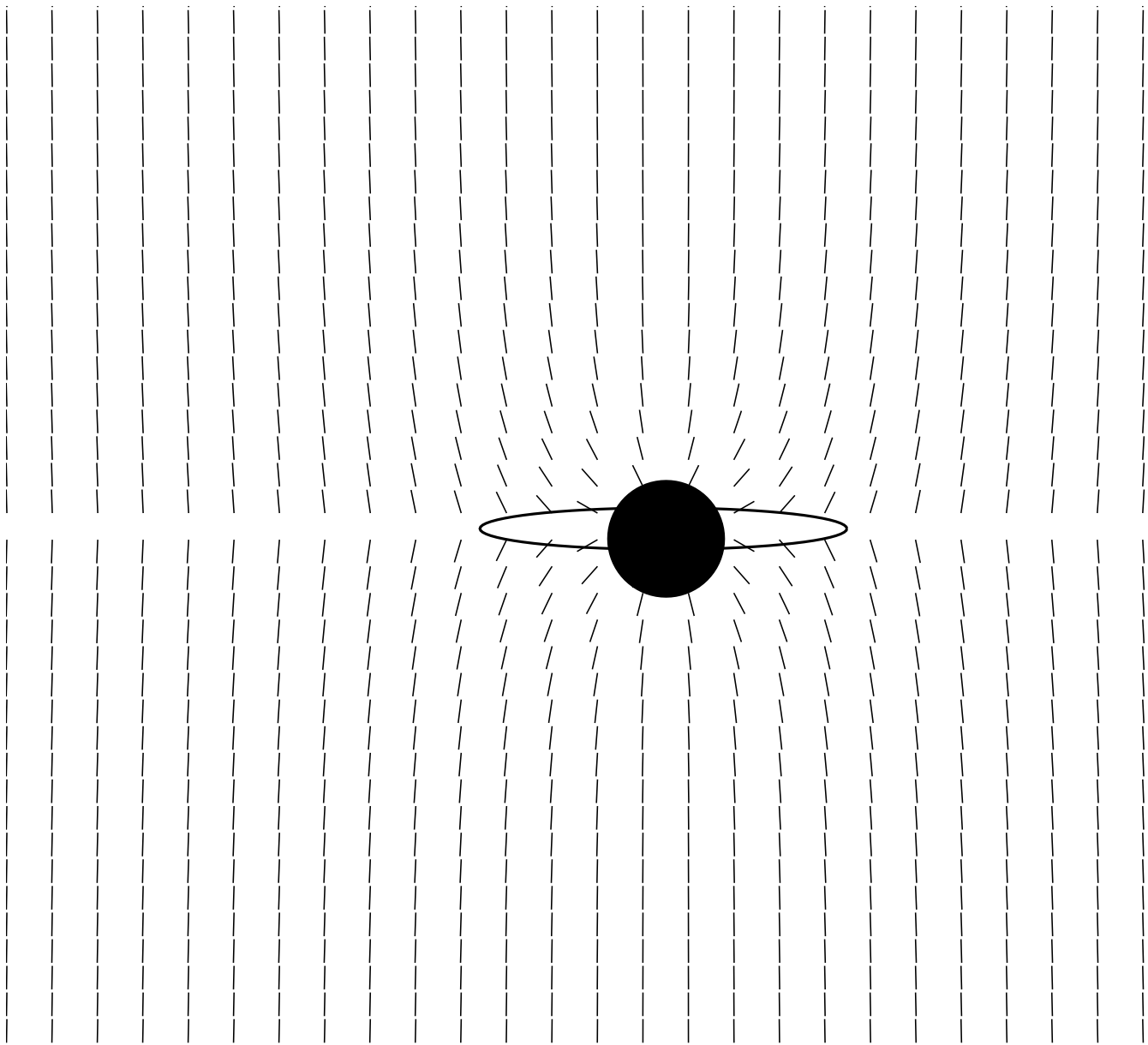
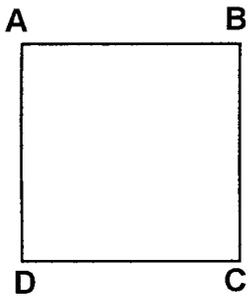
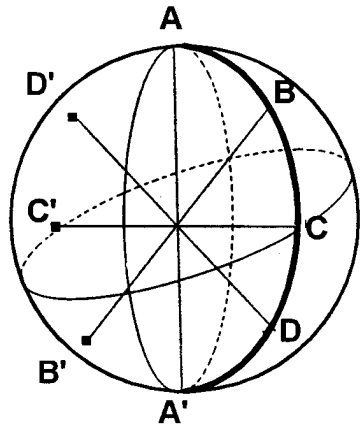
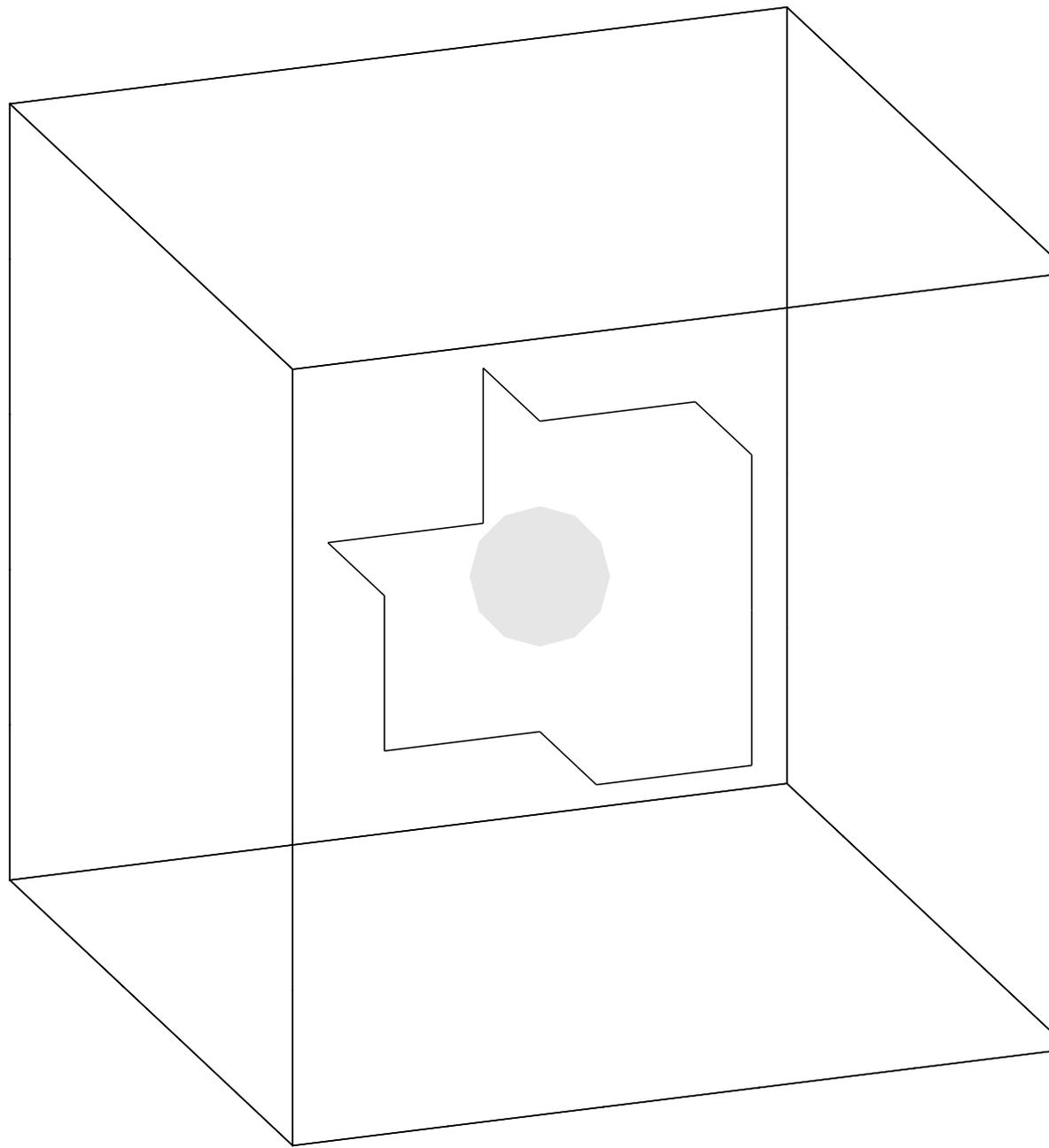
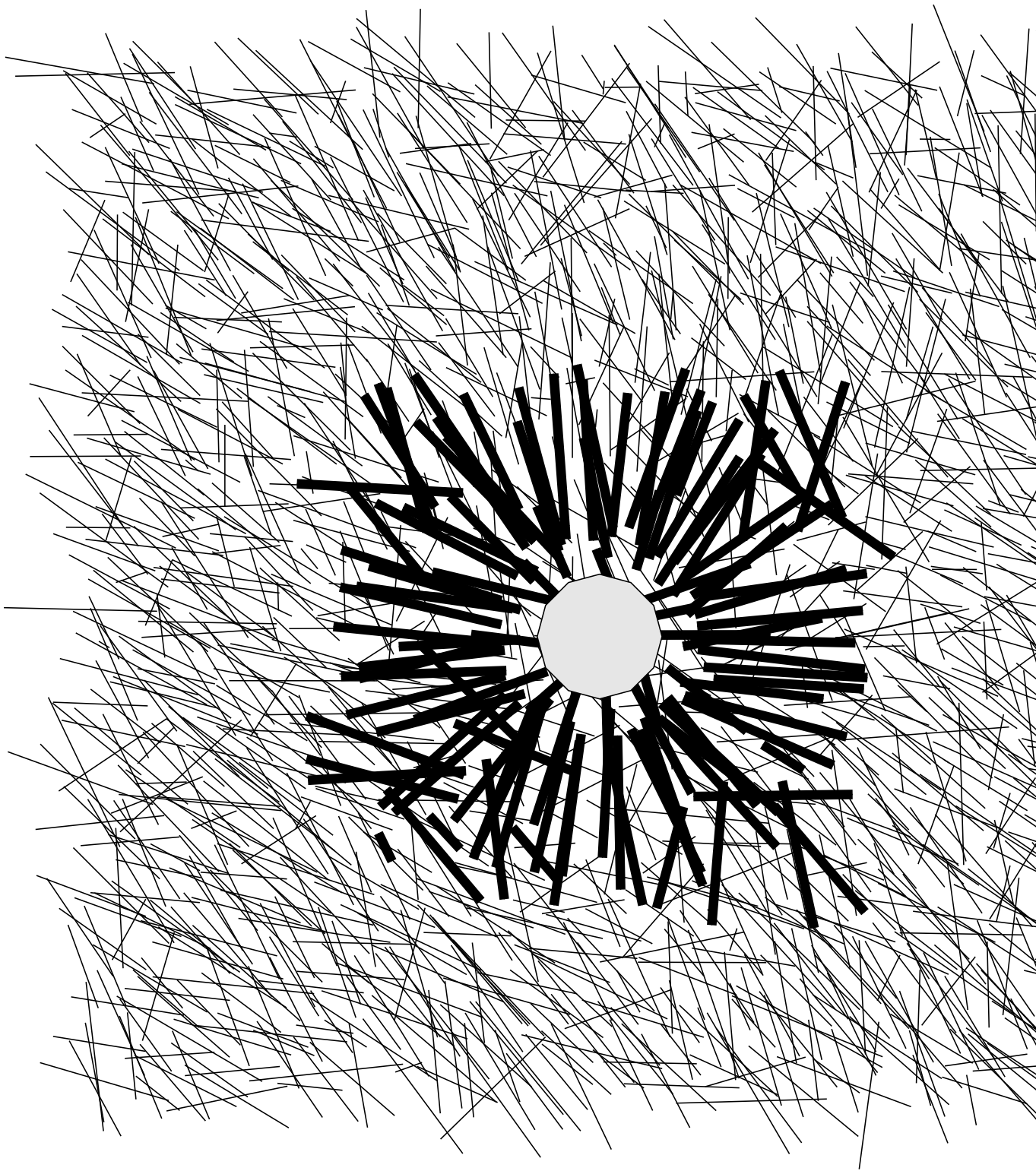
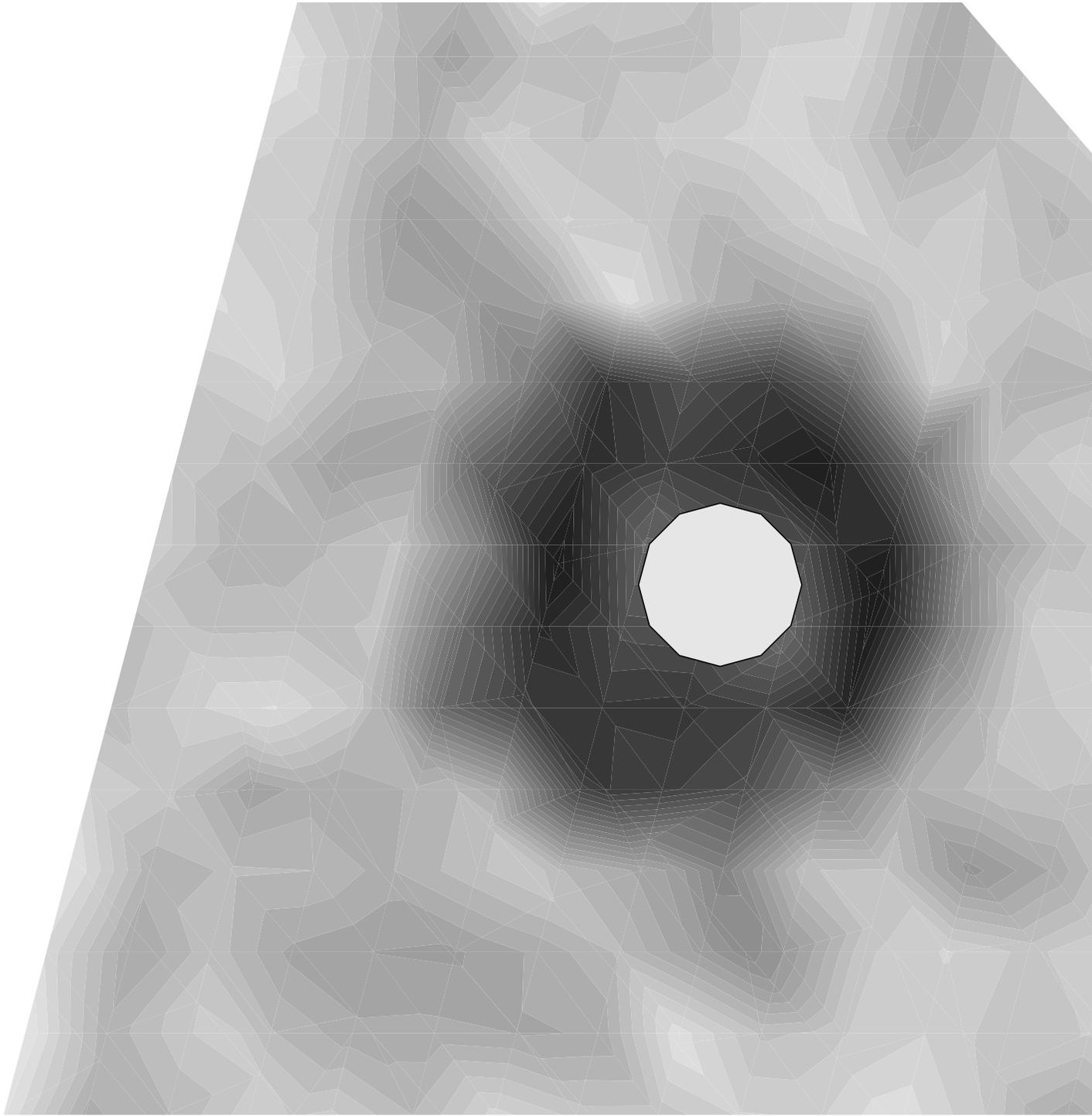
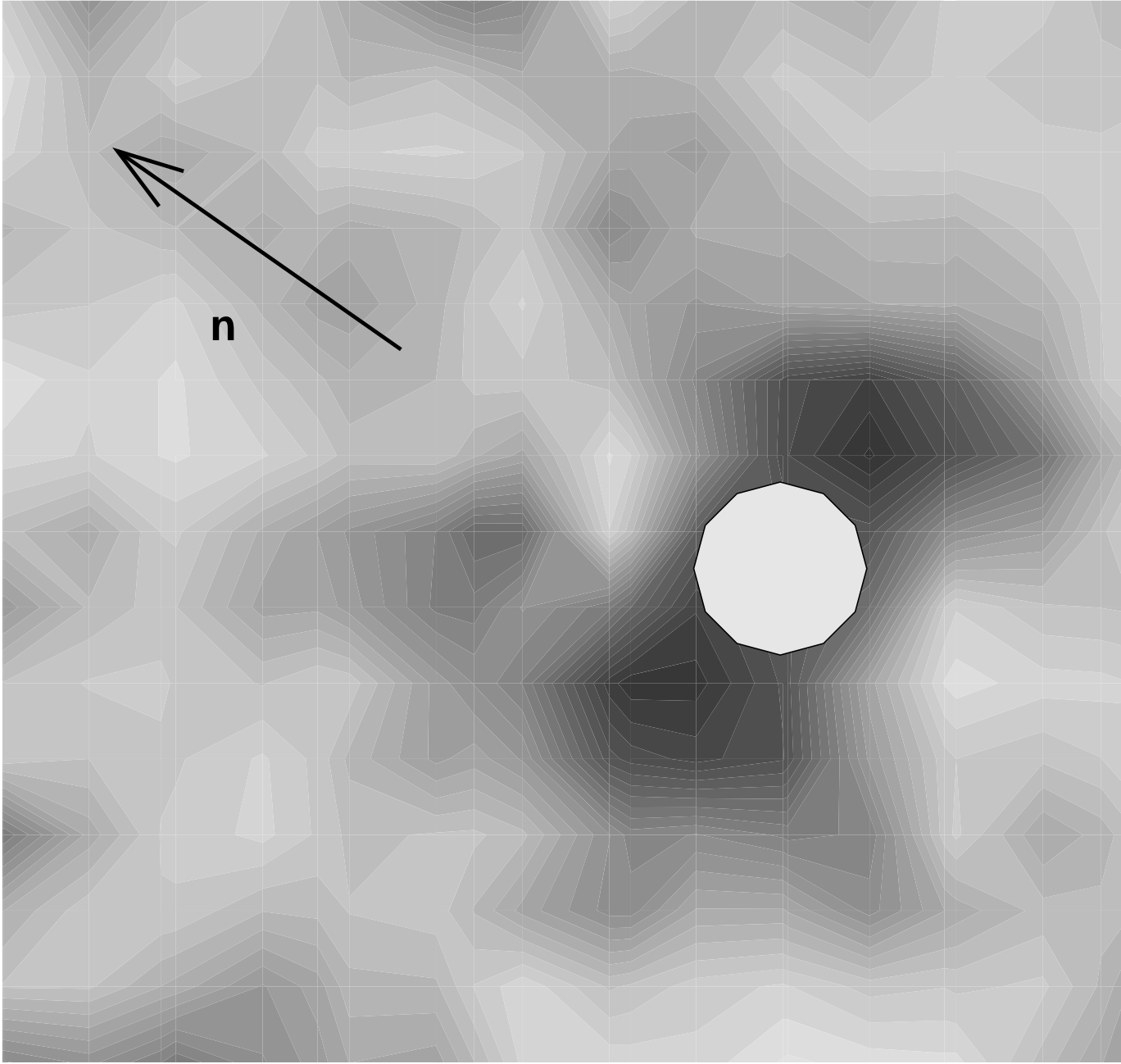
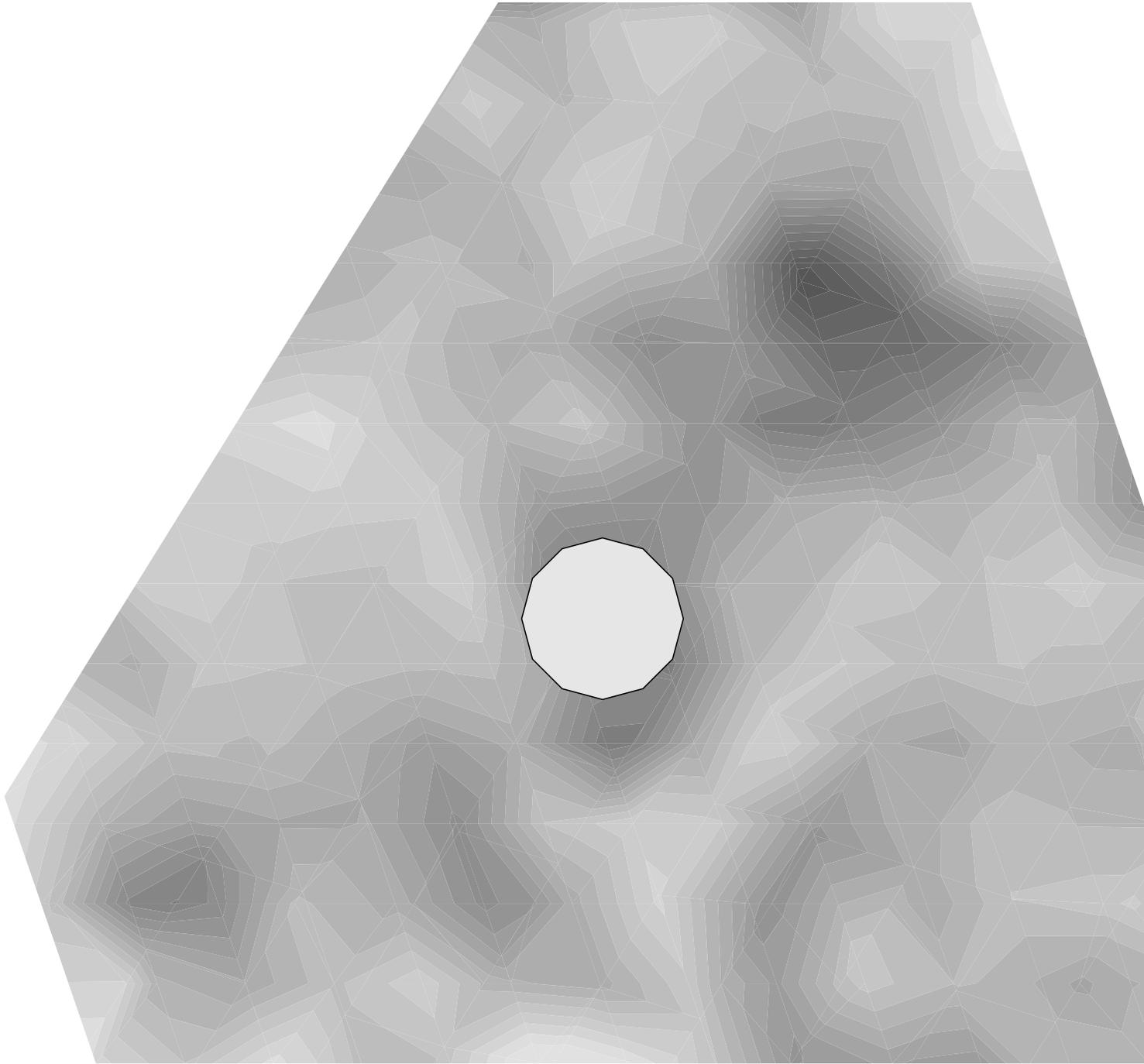
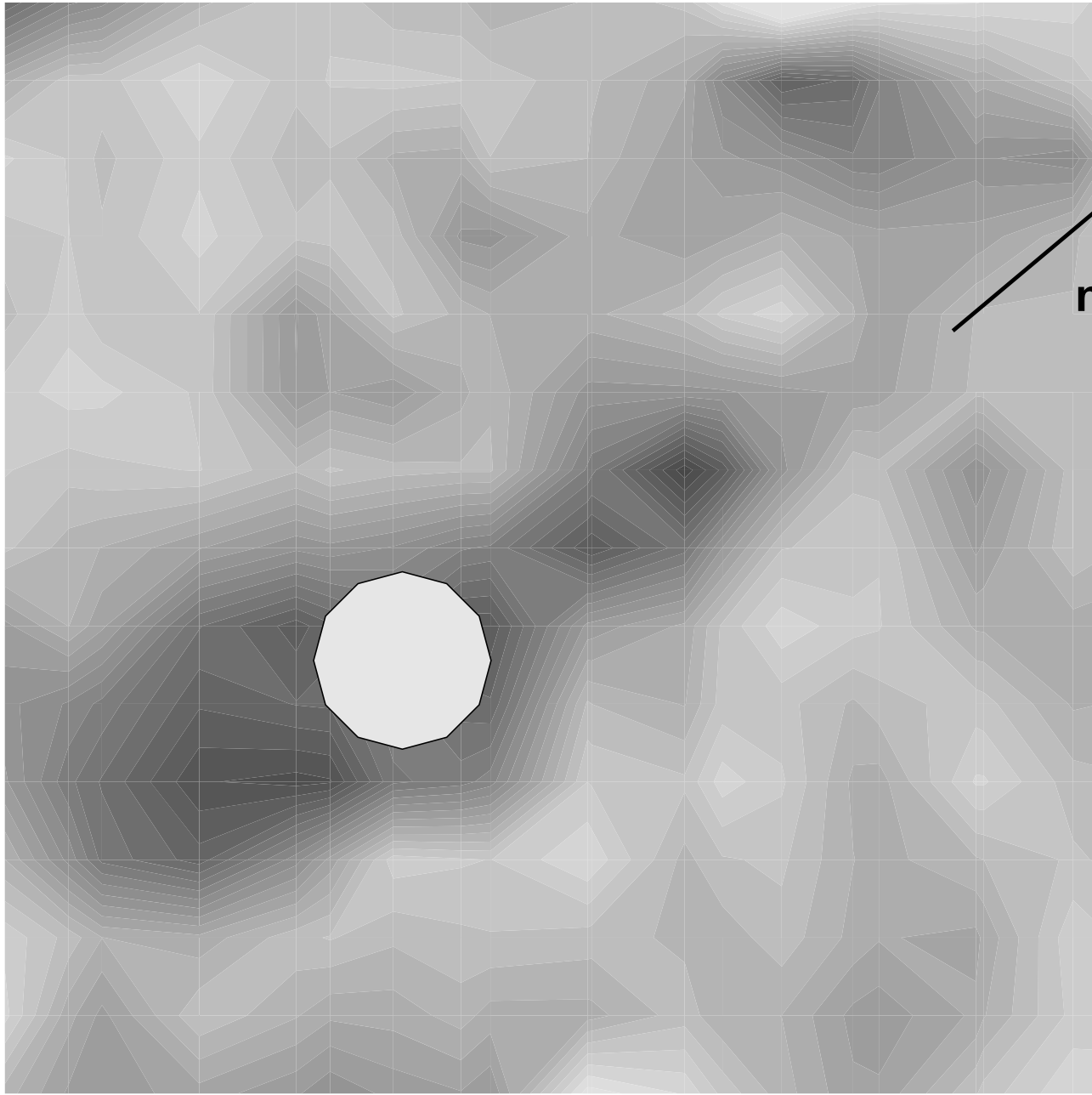
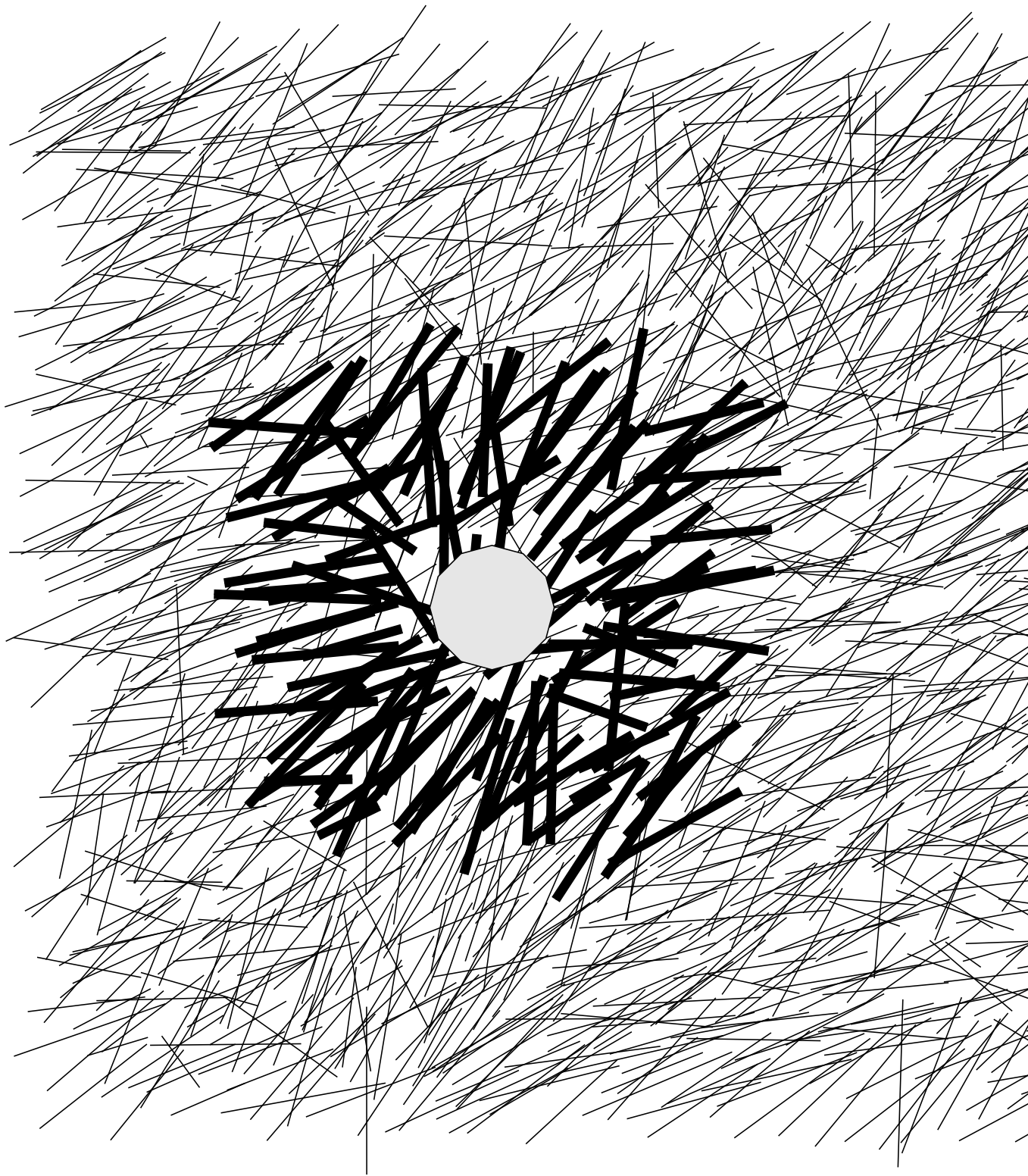
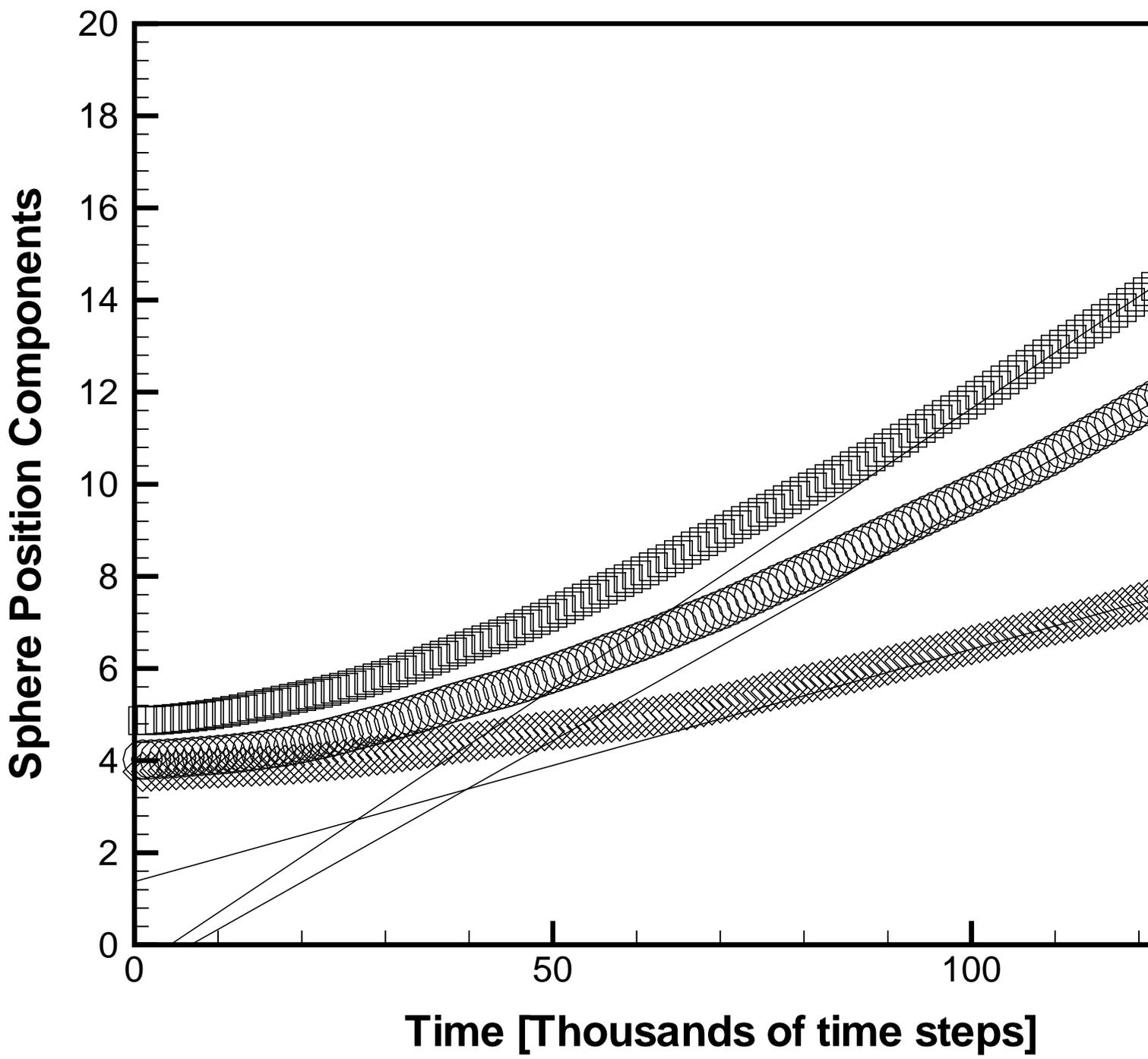
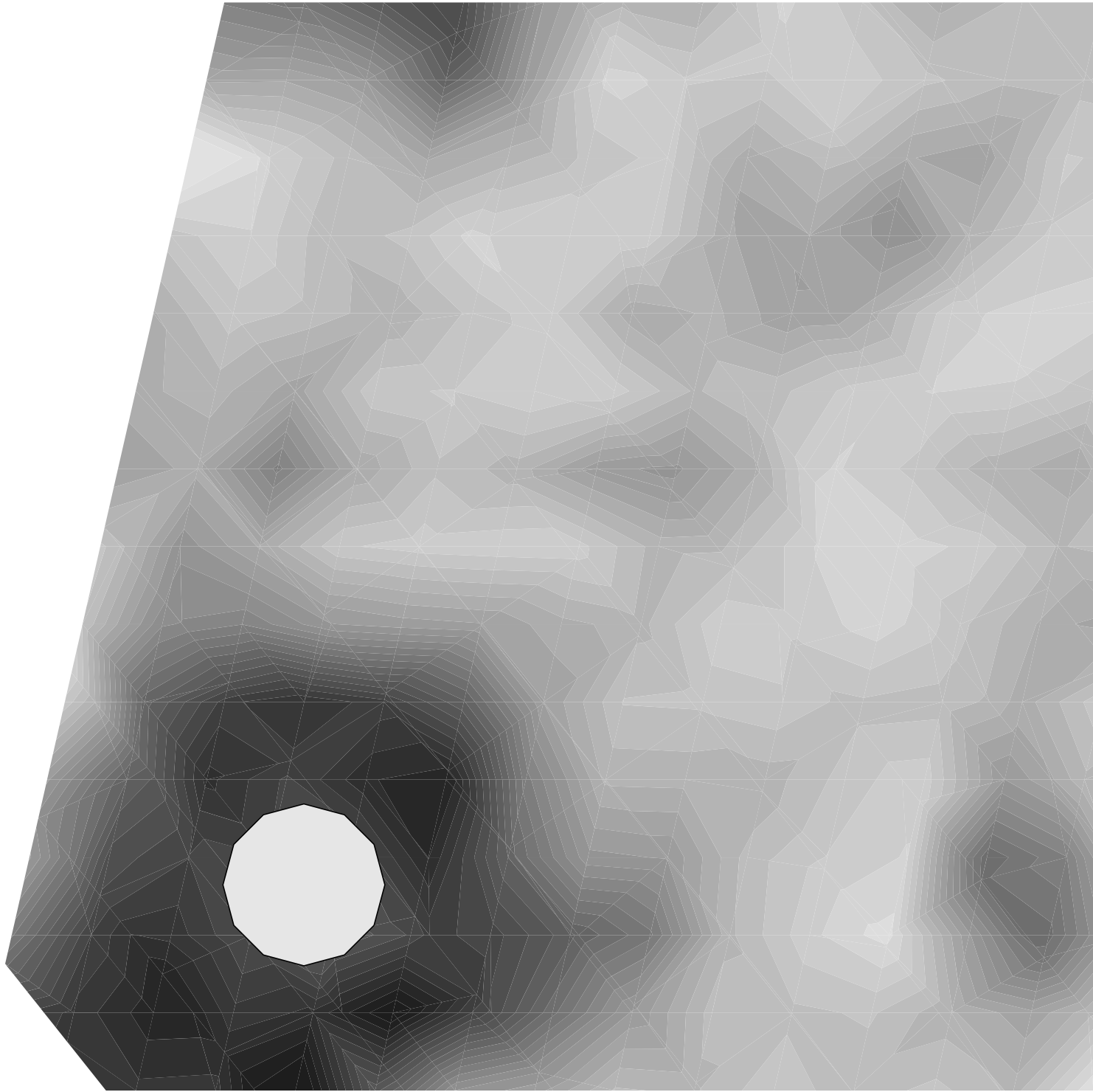
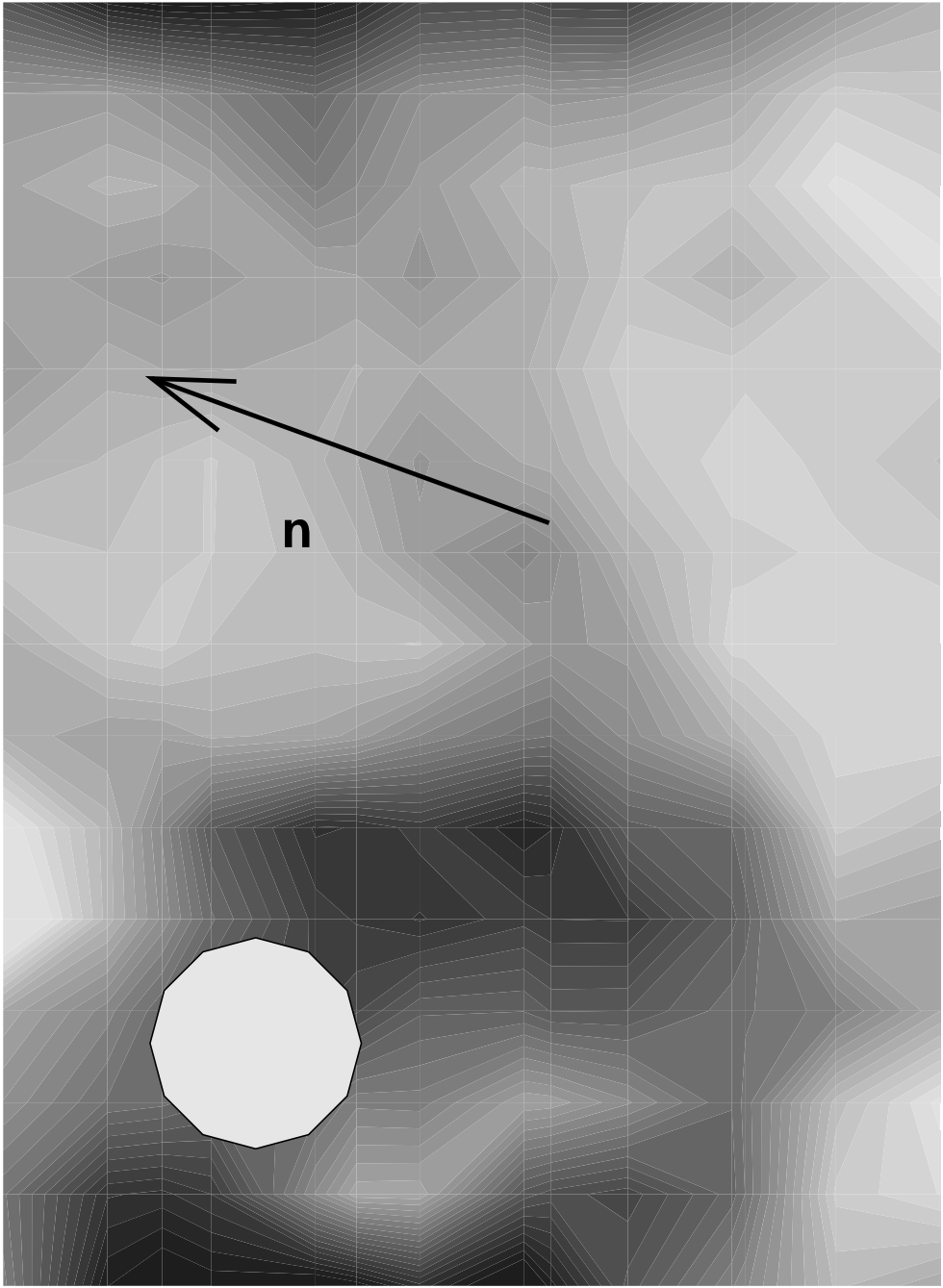
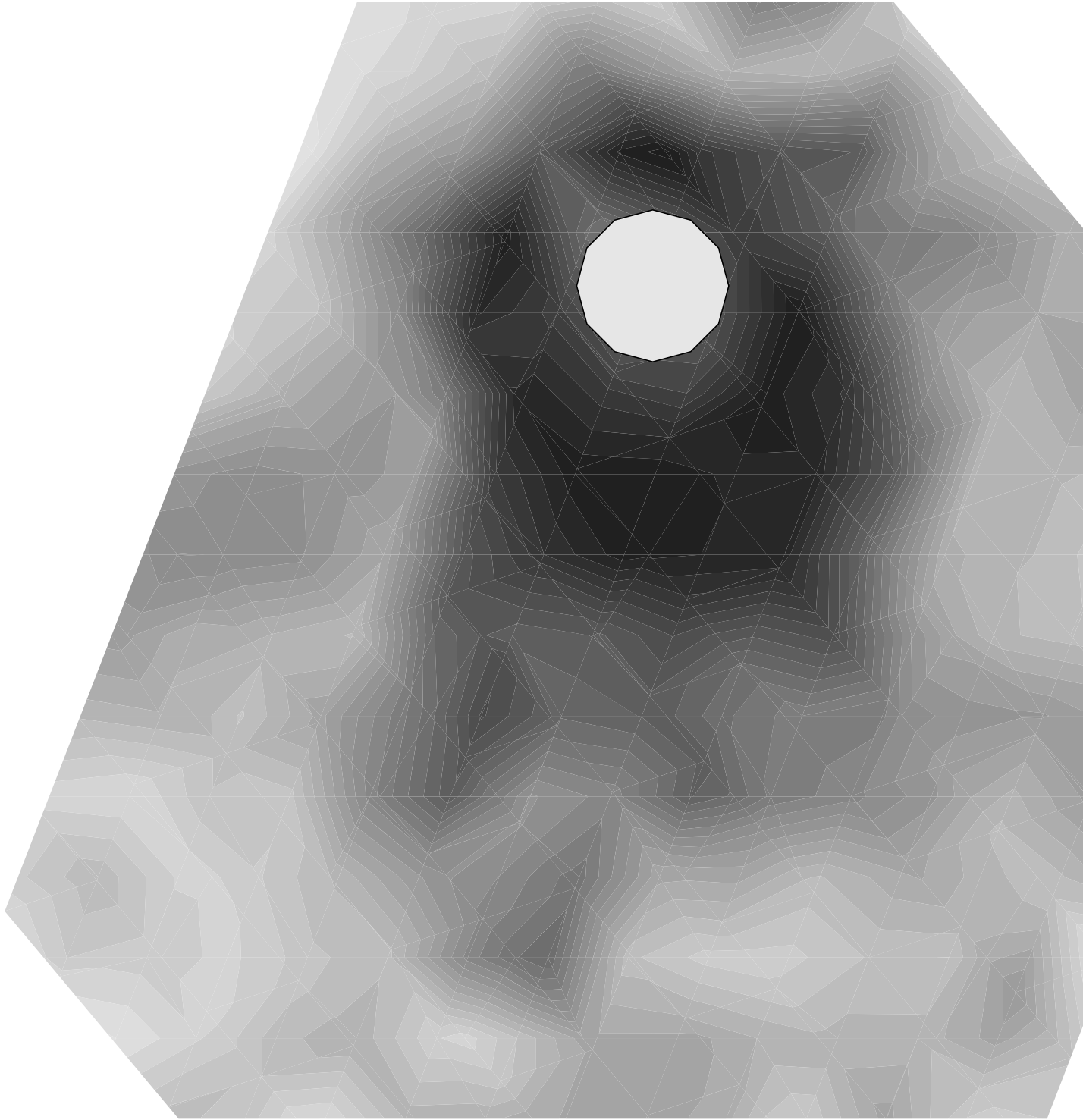
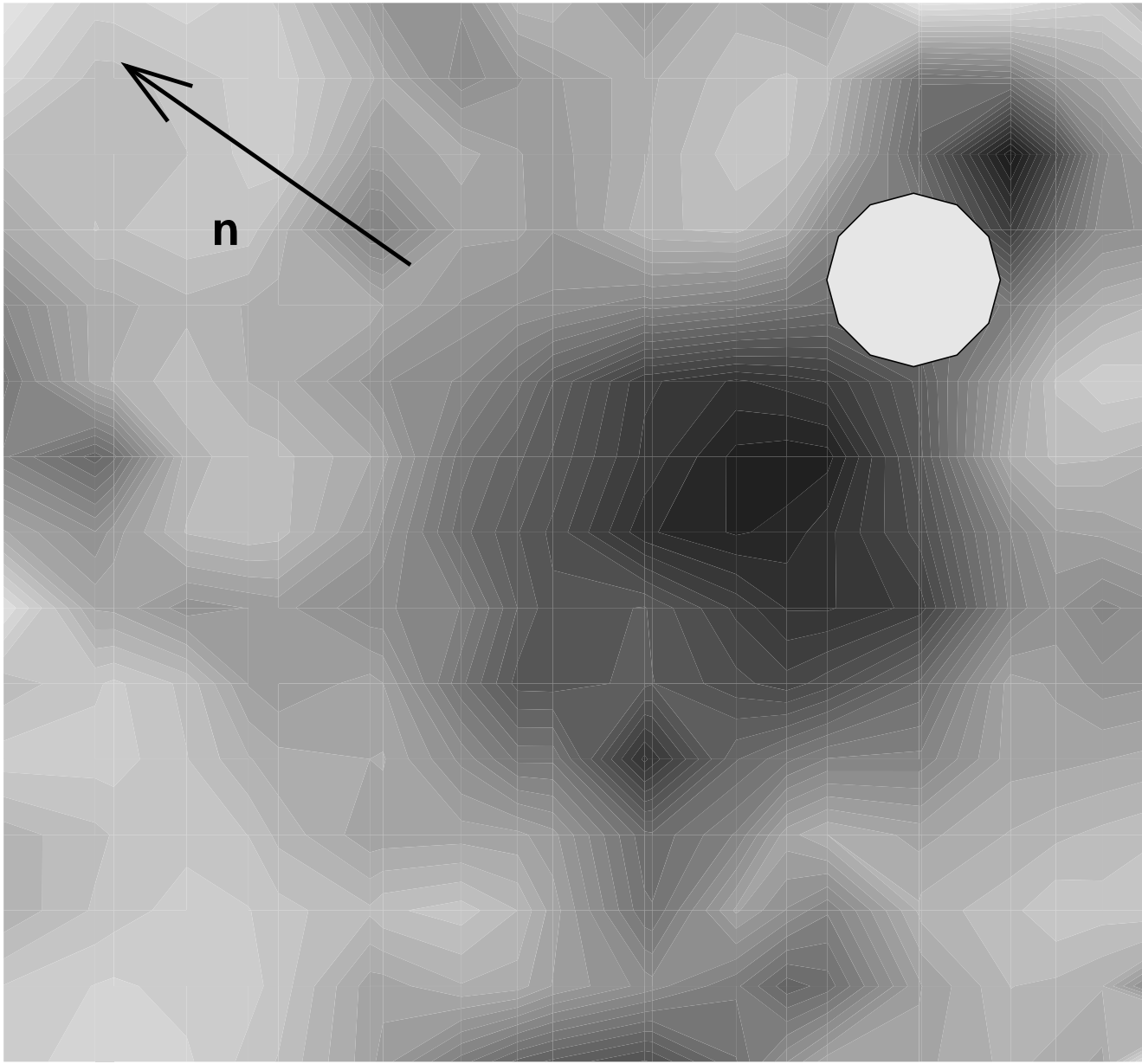
\end{document}